\documentstyle[twoside,fleqn,espcrc2]{article}

\input epsf
\epsfverbosetrue

\input epsf
\epsfverbosetrue

\title{\vspace*{-8mm}\rightline{\small  HIP--1997--53/TH}
  \vspace*{-3mm}
  \rightline{\small September 12, 1997}\vspace*{-7mm}
Comparing improved actions for SU(2)\thanks{Presented by P. Pennanen, {\tt Petrus@hip.fi}}}

\author{Petrus Pennanen\address{Helsinki Institute of Physics,
P.O. Box 9, FIN-00014 University of Helsinki, Finland}
and
Janne Peisa\address{Department of Mathematical Sciences, University of Liverpool, P.O. Box 147, Liverpool L69 3BX, United Kingdom}\thanks{\tt Janne.Peisa@helsinki.fi}}

\begin{document}

\begin{abstract}

In order to help the user in choosing the right action a performance comparison is done for seven improved actions. Six of them are Symanzik improved, one at
tree-level and two at one-loop, all with or without tadpole improvement. The 
seventh is an approximate fixed point action. Observables are
static on- and off-axis two-body potentials and four-body binding energies,
whose precision is compared when the same amount
of computer time is used by the programs.
\end{abstract}

\noindent
\maketitle

We were motivated to consider using improved actions after noting the slowness
of a four-quark flux distribution measurement code. In this case the
lattice spacing has to be small, $a\approx 0.1$ fm, to achieve sufficient 
resolution. In this work we compare actions at that scale and 
at $a\approx 0.2$ fm.

\vspace{-0.2cm}

\section{The actions}

The perturbative Symanzik approach to improvement is in this work represented
by three actions; a tree-level version with a plaquette
\setlength{\unitlength}{0.5pt}
 \begin{picture}(20,22.5)(5,11.25)
  \put(7.5,7.5){\vector(0,1){9.375}}
  \put(7.5,7.5){\line(1,0){15}}
  \put(7.5,22.5){\vector(1,0){9.375}}
  \put(7.5,7.5){\line(0,1){15}}
  \put(22.5,7.5){\vector(-1,0){9.375}}
  \put(22.5,22.5){\line(-1,0){15}}
  \put(22.5,22.5){\vector(0,-1){9.375}}
  \put(22.5,22.5){\line(0,-1){15}}
\end{picture} and a $1\times 2$ 
rectangle \begin{picture}(35,22.5)(5,11.25)
  \put(7.5,7.5){\vector(0,1){9.375}}
  \put(7.5,7.5){\line(0,1){15}}
  \put(7.5,22.5){\vector(1,0){9.375}}
  \put(7.5,22.5){\vector(1,0){24.375}}
  \put(7.5,22.5){\line(1,0){30}}
  \put(37.50,22.5){\vector(0,-1){9.375}}
  \put(37.50,22.5){\line(0,-1){15}}
  \put(37.50,7.5){\vector(-1,0){9.375}}
  \put(37.50,7.5){\vector(-1,0){24.375}}
  \put(37.50,7.5){\line(-1,0){30}}
\end{picture} \cite{lus:85a} (abbreviation: S) and two one-loop actions, one with a $1^3$ parallelogram (x,y,z,-x,-y,-z) \begin{picture}(32,22.5)(5,11.25)
  \put(7.5,7.5){\vector(0,1){9.375}}
  \put(7.5,7.5){\line(0,1){15}}
  \put(7.5,22.5){\vector(2,1){7.5}}
  \put(7.5,22.5){\line(2,1){11.25}}
  \put(18.9,28.2){\vector(1,0){9.375}}
  \put(18.9,28.2){\line(1,0){15}}
  \put(33.9,28.2){\vector(0,-1){9.375}}
  \put(33.9,28.2){\line(0,-1){15}}
  \put(33.9,13.2){\vector(-2,-1){7.5}}
  \put(33.9,13.2){\line(-2,-1){11.25}}
  \put(22.5,7.5){\vector(-1,0){9.375}}
  \put(22.5,7.5){\line(-1,0){15}}
\end{picture} \cite{lus:85a,wei:84} (S1) and the other with both parallelogram
and a $2\times 2$ large square \begin{picture}(35,35)(5,11.25)
  \put(7.5,7.5){\vector(0,1){9.375}}
  \put(7.5,7.5){\vector(0,1){24.375}}
  \put(7.5,7.5){\line(0,1){30}}
  \put(7.5,37.5){\vector(1,0){9.375}}
  \put(7.5,37.5){\vector(1,0){24.375}}
  \put(7.5,37.5){\line(1,0){30}}
  \put(37.50,37.5){\vector(0,-1){9.375}}
  \put(37.50,37.5){\vector(0,-1){24.375}}
  \put(37.50,37.5){\line(0,-1){30}}
  \put(37.50,7.5){\vector(-1,0){9.375}}
  \put(37.50,7.5){\vector(-1,0){24.375}}
  \put(37.50,7.5){\line(-1,0){30}}
\end{picture} \cite{sni:96} (S1S) as additional operators. Of these also the 
tadpole improved (TI) versions (STI, S1TI, S1STI) are considered. 
TI for the S1 action follows Ref. \cite{alf:95} using results for SU(2) in 
Ref. \cite{wei:84}. 

The non-perturbative approach to improvement is 
represented by a truncated fixed point action (FP)
which includes first to fourth powers of the plaquette and the 
parallelogram \cite{deg:96}.

\vspace{-0.2cm}

\section{The task}

The measurements consist of static two-quark potentials for
$R=1,\ldots,6$ on-axis, $R=(1,1),(2,1),\ldots,(3,3)$ off-axis and
the binding energies of four quarks at the corners of a regular tetrahedron, 
the cube surrounding it having sides of length $R=1,2,3$. Here binding 
energies mean $E_4 - 2V_2$, where $E_4$ is
the energy of four quarks and $2V_2$ the energy of the lowest-lying
two-body pairing -- see \cite{glpm:96,pen:96b}.

In order to separate the ground state a variational basis of 
fuzzing levels 13 and 2 is used.  An update step consisted of four
overrelaxations and one heatbath sweep, except for the FP
case for which the latter was replaced by ten Metropolis sweeps. 
Table \ref{tbeta} shows the $\beta$ values and corresponding scales used for
the comparison. Scales 
were set by fitting plateau two-body potentials at $R=2,\ldots,6$ with the 
continuum parameterization $V_{\rm fit}=-e/R+b_S R+V_0$
 and using Sommer's equation $r_0^2 F(r_0)=c$ with $c=2.44$,
corresponding to $r_0\approx 0.66$ fm. These scales agree with the determination using $\sqrt{b_S}=440$ MeV. The plateau was taken to be reached
when the difference of potentials at $T+1$ and $T$ was smaller than the 
bootstrap error on this difference. 

\begin{table}[hbt]
\vspace{-1.5cm}
\begin{tabular}{lcccc}
       & \multicolumn{2}{c}{$a\approx 0.1$ fm}  & \multicolumn{2}{c}{$a\approx 0.2$ fm}  \\ \hline
Action &$\beta$ & $a$ [fm] & $\beta$ & $a$ [fm]\\ \hline
W      &  2.45  & 0.101(3) & 2.23 & 0.204(3)     \\
S      &  1.86  & 0.092(5) &  1.63 & 0.197(5)  \\
STI     &  1.98  & 0.097(2) &  1.73 & 0.202(4)   \\
S1     &  3.45  & 0.096(4) &  3.03 & 0.198(2) \\
S1TI    &  3.5   & 0.097(3) & 3.065 & 0.211(7)   \\
S1S    &  3.65  & 0.097(5) &  3.23  & 0.202(3)  \\
S1STI   &  3.75  & 0.094(4) &  3.33 & 0.196(4)  \\
FP     &  1.69  & 0.101(3) &  1.502 & 0.213(5)  \\ 
\end{tabular}
\caption{
\label{tbeta}}
\vspace{-0.5cm}
\end{table}

The runs at $a\approx 0.1$ fm ($a\approx 0.2$ fm) on a $16^4$ ($12^4$) lattice 
consisted of 1000 (2500)
measurements and were performed on SGI R10000 workstations. In the following, unless noted otherwise, we 
take a 
subset of these corresponding to the same amount of total CPU 
time consumed.
 
\section{Results}

\underline{CPU time:} Table \ref{tcpu} shows the CPU time per update and per 
update+measurement relative to the Wilson
action (W) -- the 
measurements should take the same amount of time. 
Variations in consumption for two different actions using the same 
operators (one of them TI) reflect mostly the systematic errors 
in our time measurement.
These are probably due to variations of CPU load
and free memory. 

The autocorrelation times of the plaquette average do not seem to depend on the
size of the operators in the action. 

\begin{table}[hbt]
\vspace{-0.6cm}
\begin{tabular}{lcccc} 
       & \multicolumn{2}{c}{$a\approx 0.1$ fm}  & \multicolumn{2}{c}{$a\approx 0.2$ fm}  \\ \hline
       &  upd./tot. & a.c. & upd./tot. & a.c. \\ \hline
W      &  1/1   & 1.9(4) & 1/1   & 3.5(6) \\
S      &  3.3/1.2 & 2.1(5) & 3.6/1.4 & 4.5(9) \\
STI     &  2.9/1.1 & 1.3(2) & 3.5/1.4 & 3.3(6) \\
S1     &  7.9/1.5 & 1.6(4) & 9.3/2.2 & 2.5(4)  \\
S1TI    &  8.0/1.5 & 1.9(5) & 9.6/2.3 & 2.4(4)  \\
S1S    &  11.2/1.8 & 2.3(8) & 12.0/2.7 & 2.7(5) \\
S1STI   &  12.4/1.8 & 1.7(4) & 10.4/2.4 & 4.5(9)  \\
FP     &  12.4/1.8 & 0.5(5) & 10.4/2.4 & 3.0(5) \\ 
\end{tabular}
\caption{CPU time and autocorrelation. \label{tcpu}}
\vspace{-0.4cm}
\end{table}

\underline{Plateaux:} 
Actions which violate reflection positivity with negative eigenvalues of the 
Hamiltonian usually have a local maximum 
in the $M_{\rm eff}$ plot before a plateau is reached. 
When comparing the behaviour with the same number of measurements used, TI 
improves the plateau after this 'bump', most notably for 
the STI case at $a\approx 0.1$ fm,
which means that less correlators are needed. This can be used to save CPU 
time. Wilson and FP actions also have good plateaux and do not violate 
reflection positivity.

\pagebreak
\underline{Statistical errors and rotational variance:}
Table \ref{terr} shows the average relative errors, which are statistical for 
all other observables except for the off-axis two-body potentials (Fig. 1), 
for which 
the deviation $|V_{\rm meas}-V_{\rm fit}|/V_{\rm fit}$ from the value given 
by the on-axis fit is also shown. The best improved actions are seen 
to have less rotational variance at $a\approx 0.2$ fm than the Wilson action
at $a\approx 0.1$ fm.

\begin{table}[hbt]
\vspace{-0.8cm}
\begin{tabular}{lccccc}
       & \multicolumn{3}{c}{$a\approx 0.1$ fm}  & \multicolumn{2}{c}{$a\approx 0.2$ fm}  \\ \hline
       & on- & off-axis & 4q & on- & off-axis \\ \hline
W      &  {\bf 0.37} & 1.39(9) & 4.8 & {\bf 0.29} & 2.35(9)   \\
S      &  0.54 & 0.72(16) & 4.4 & 0.66 & 1.23(13) \\
STI     &  {\bf 0.36} & {\bf 0.58(9)}  & {\bf 2.2} & 0.41 & 0.93(18) \\
S1     &  0.56 & 0.66(14) & 5.3 & 0.59 & 0.97(12) \\
S1TI    &  0.49 & 0.68(11) & 4.9 & 0.37 & {\bf 0.46(17)} \\
S1S    &  0.86 & 0.83(22) & 6.3 & 0.5 & 1.72(20) \\
S1STI   &  0.90 & 0.67(23) & {\bf 3.1} & 0.9 & 0.83(16) \\
FP     &  {\bf 0.43} & {\bf 0.49(10)} & {\bf 3.1} & 0.54 & 1.77(46)  \\ 
\end{tabular}
\caption{Average errors ($\times 100$) in two-body potentials and four-body binding energies.\label{terr}}
\vspace{-0.8cm}
\end{table}

\begin{figure}[p]
\vspace{-1cm}
\hspace{0cm}\epsfysize=150pt\epsfxsize=220pt\epsfbox{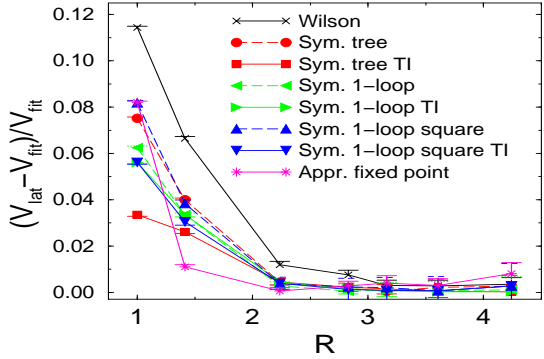}
\vspace{-1.1cm}
\caption{Rotational variance and statistical error for off-axis potentials at $a\approx 0.1$ fm. \label{foff}}
\end{figure}

\underline{Scaling:} Figs. 2 and 3 show the differences of measured potentials from 
a fit to Wilson action data at $\beta=2.85$ \cite{boo:93}, 
corresponding to $a\approx 0.027$ fm. 
In table \ref{tcd2} the averages 
of these differences are shown. None of the improved actions at $a\approx 0.2$ fm 
can be seen to scale as well as the Wilson action at $a\approx 0.1$ fm. 

\begin{table}[hbt]
\vspace{-0.8cm}
\begin{tabular}{lcc} 
       & $a\approx 0.1$ fm                & $a\approx 0.2$ fm  \\ \hline
W      &  0.020(4)  & 0.052(7) \\
S      &  0.015(7) & 0.063(12) \\
STI    &  {\bf 0.012(4)} & {\bf 0.042(11)} \\
S1     &  0.022(7) & 0.045(10) \\
S1TI   &  {\bf 0.009(6)} & {\bf 0.039(11)} \\
S1S    &  0.027(11) & 0.047(11) \\
S1STI  &  {\bf 0.013(10)} & {\bf 0.043(15)} \\
FP     &  0.027(6) & 0.064(13) \\
\end{tabular}
\caption{\label{tcd2}}
\vspace{-1.4cm}
\end{table}

\begin{figure}[p]
\vspace{-1.2cm}
\hspace{0cm}\epsfysize=150pt\epsfxsize=220pt\epsfbox{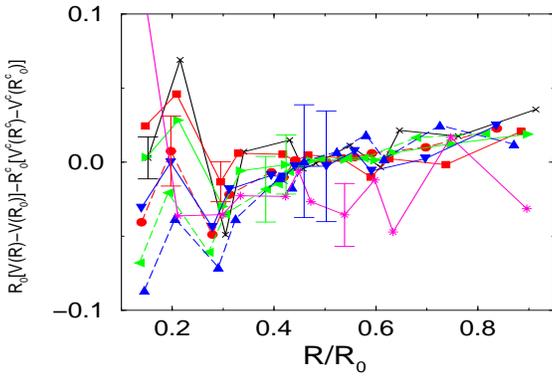}
\vspace{-1.1cm}
\caption{Difference of $a\approx 0.1$ fm potentials from those at $a=0.027$ fm. For each action one error bar is shown. \label{fdiff}}
\end{figure}

\begin{figure}[p]
\vspace{-1.2cm}
\hspace{0cm}\epsfysize=150pt\epsfxsize=220pt\epsfbox{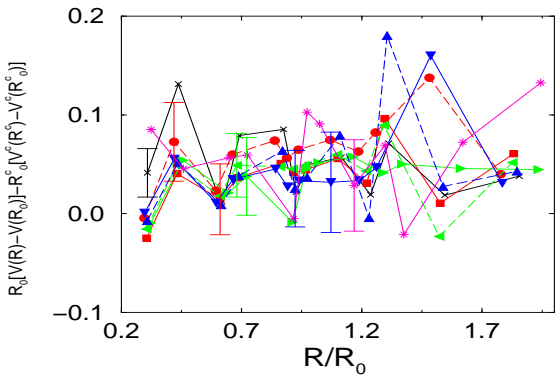}
\vspace{-1.1cm}
\caption{As Fig. 2 but for $a\approx 0.2$ fm\label{fdiff}}
\end{figure}

\underline{Self-energies:}
The lattice self-energies of the quarks given by the on-axis fit vary up to 
40 \% between different actions. In lowest-order perturbation theory this is 
due to the different lattice one-gluon exchange operators.

\pagebreak

\section{Discussion}

For two-body potentials rotational invariance is improved significantly, while
statistical errors are not --
with the same amount of computer time used the Wilson action has the smallest
statistical errors. The improvement in scaling seems to be small. 
For Symanzik improved actions TI works; statistical
errors, rotational variance and scaling violations are reduced. 
Our only representative of truncated FP actions performs well at 
$a\approx 0.1$ fm, but has problems at the coarser $a$. 

When using an improved action further improvement can be achieved with
improved operators, which can also be essential for correct physics e.g. in 
lattice sum rules. Technical difficulties associated with improved operators 
include the separation of field components, for which planar actions (in this 
work S, STI) are easier. Other means of improvement include anisotropic 
lattices, lookup tables for loop collection, cache optimization and tuning
the fuzzing parameters. Some of the latter three can be quite easily 
implemented with possibly a significant time-saving effect.

{\bf Acknowledgement:}
We thank A.M. Green and C. Michael for support and discussions,
the Finnish Academy (P.P) and Magnus Ehrnrooth Foundation for funding and 
the Helsinki Institute of Physics for hospitality and computing facilities.

\end{document}